\documentclass[prb,twocolumn,showpacs,floatfix]{revtex4}
\usepackage{graphicx}
\usepackage{dcolumn}
\usepackage{bm}
\begin{document}
\title{Magnetoelastic effects in Jahn-Teller distorted CrF$_2$ and CuF$_2$ studied by  neutron powder diffraction}
\author{Tapan Chatterji and  Thomas C. Hansen }
\address{Institut Laue-Langevin, B.P. 156, 38042 Grenoble Cedex 9, France\\}
\date{\today}

\begin{abstract}
We have studied the temperature dependence of crystal and magnetic structures of the Jahn-Teller distorted transition metal difluorides CrF$_2$ and CuF$_2$ by neutron powder diffraction in the temperature range $2-280$ K. The lattice parameters and the unit cell volume show magnetoelastic effects below the N\'eel temperature. The lattice strain due to the magnetostriction effect couples with the square of the order parameter of the antiferromagnetic phase transition. We also investigated the temperature dependence of the Jahn-Teller distortion which does not show any significant effect at the antiferromagnetic phase transition but increases linearly with increasing temperature for CrF$_2$ and remains almost independent of temperature in CuF$_2$. The magnitude of magnetovolume effect seems to increase with the low temperature saturated magnetic moment of the transition metal ions but the correlation is not at all perfect.
\end{abstract}
\pacs{75.25.+z}
\maketitle
\section{Introduction}
The coupling between the spin and lattice degrees of freedom has turned out to be a central topic in condensed matter physics.  At the early stage of condensed matter research, though knowing fully well its importance, one had the tendency to neglect this coupling deliberately in order to avoid complications. Recently, however, thanks to both better experimental as well as theoretical and computational capabilities, the coupling between  the spin and lattice degrees of freedom have attracted renewed interest. The dependences on distance of the exchange integral, the spin-orbit and the dipole-dipole interactions imply that the spin system of a ferromagnetic or antiferromagnetic crystal is coupled to the ionic displacements. The static portion of this interaction results in a shift in the equilibrium ionic positions relative to the case with zero magnetoeastic coupling, with resultant shifts of phonon and magnon spectra. The dynamic portion of the interaction gives magnon-phonon scattering. The simplest aspect of the static interaction is the spontaneous magnetostriction, or change in the crystal dimensions below the ordering temperature in zero magnetic field. The magnetostriction in applied magnetic field is called forced magnetostriction, which will not be considered here.  The crystal symmetry dictates of course the magnetoelastic modes. Manetoelastic coupling has been treated extensively by Callen and Callen \cite{callen63,callen65}.  
This coupling has been investigated in potentially important electronic materials, such as colossal magnetoresistive (CMR) manganites and multiferroic materials. One aspect of this coupling is the spontaneous static magnetoelastic effect which is sometimes called magneto-striction or exchange-striction. The lattice tends to respond  by distorting itself as the magnetic order sets in. The distortion involves the lattice and sometimes positional parameters with or without a change in symmetry. However, the effect being very small it can only be detected and investigated by high resolution synchrotron X-ray and neutron diffraction techniques. It is useful to have access to both techniques because although X-ray diffraction has usually better momentum space resolution suited for studying small changes in the lattice parameters and symmetry, the neutron diffraction is better suited to study the magnetic structures and small deviations in atom coordinates of light atoms in presence of heavier ones. Also it is necessary to measure the diffraction patterns using very fine temperature steps and for this a high intensity X-ray or neutron beam is useful in order to complete the study in a finite time. 

The exchange striction has been investigated quite intensively in ferromagnets due to their industrial applications and has been reviewed by several authors \cite{mayergoyz00,tremolet93,morin90,andreev95,wassermann90,clark80}. The magnetoelastic effects in antiferromagnets have been relatively less investigated.  However the antiferromagnetic transition metal oxides and chalcogenides have been investigated \cite{morosin70} throughly. Also the magnetoelastic effects in rare earth metals and compounds have been reported \cite{doerr05,lindbaum02}. We recently reported the results of our neutron diffraction investigations \cite{chatterji08,chatterji09a,chatterji10a,chatterji10b} of the magnetoelastic effects in antiferromagnetic manganites LaMnO$_3$, NdMnO$_3$ and  also the antiferromagnetic transition metal difluorides MnF$_2$, FeF$_2$, CoF$_2$ and NiF$2$.

\begin{figure}
\resizebox{0.5\textwidth}{!}{\includegraphics{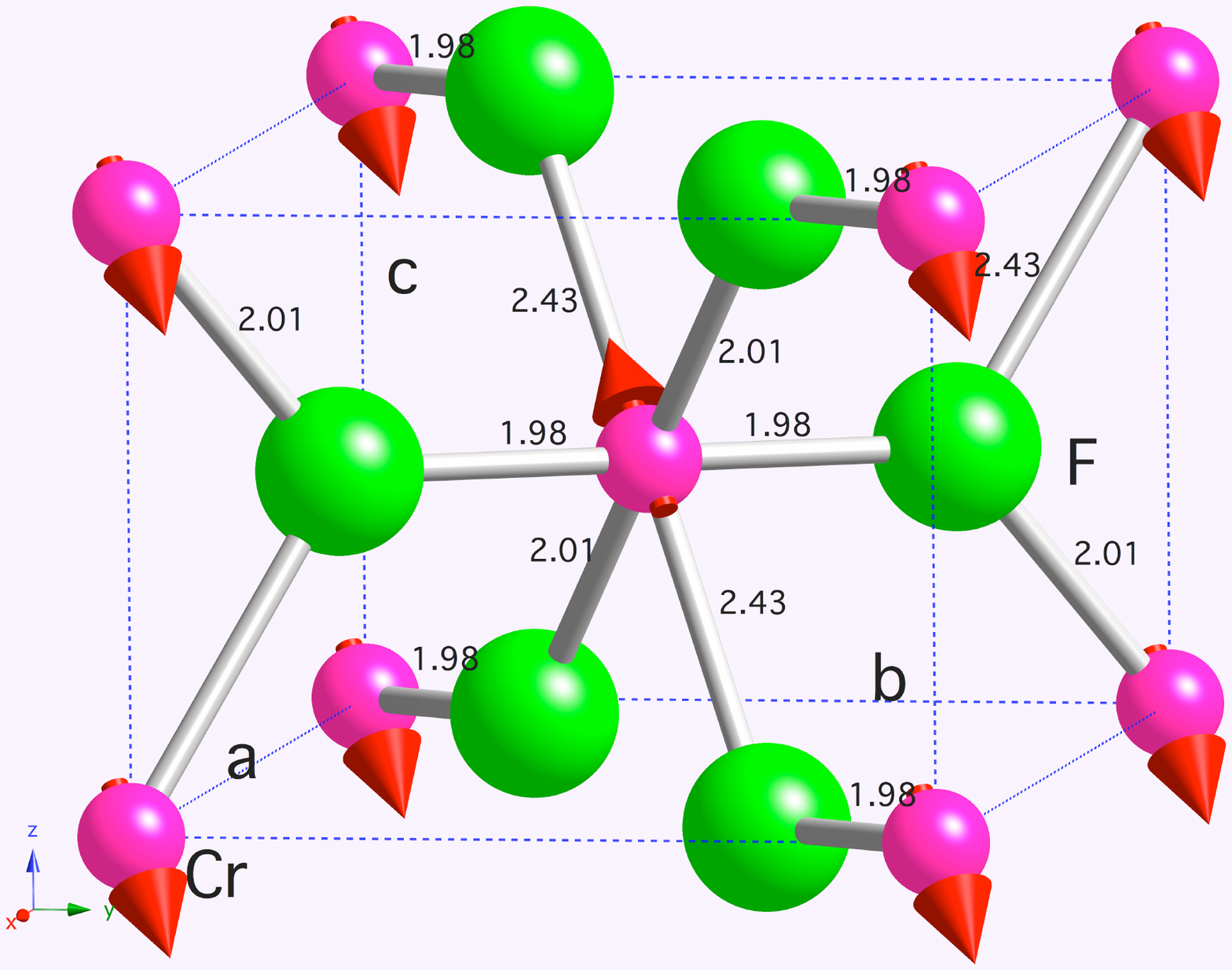}}
\caption {The distorted rutile-type monoclinic crystal structure adopted by $3d$ 
transition metal diflurides MF$_2$ (M = Cr,Cu). The red circles represent  the transition metal M ion and the green circles represent the F ions.The Jahn-Teller M ions are surrounded by F ions to produce distorted octahedra. The crystallographic axes and the M-F bond distances correspondinfg to CrF$_2$ only are shown. The magnetic structure of the ordered phase below the N\'eel temperature is indicated by showing the magnetic moment directions by the red arrows.
}
 
\label{structure}
\end{figure}

\begin{figure}
\resizebox{0.5\textwidth}{!}{\includegraphics{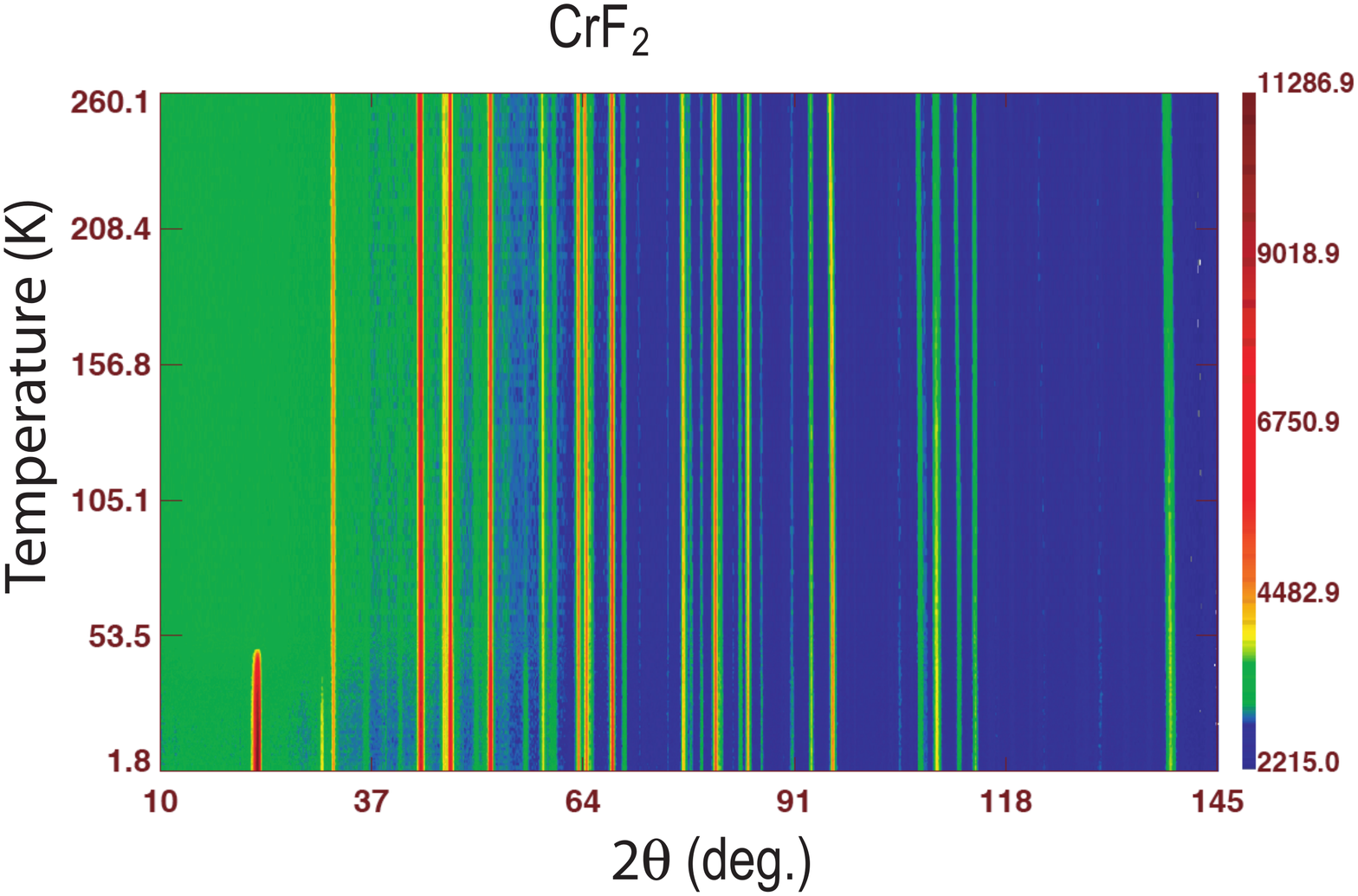}}
\caption {(Color online) Temperature variation of the diffraction diagram of CrF$_2$. The peak at about $2\theta \approx 20$ deg. is the $100/010$ magnetic reflection whose intensity decreases continuously with increasing temperature and becomes zero at $T_N \approx 50$ K. } 
\label{lampCr}
\end{figure}

\begin{figure}
\resizebox{0.5\textwidth}{!}{\includegraphics{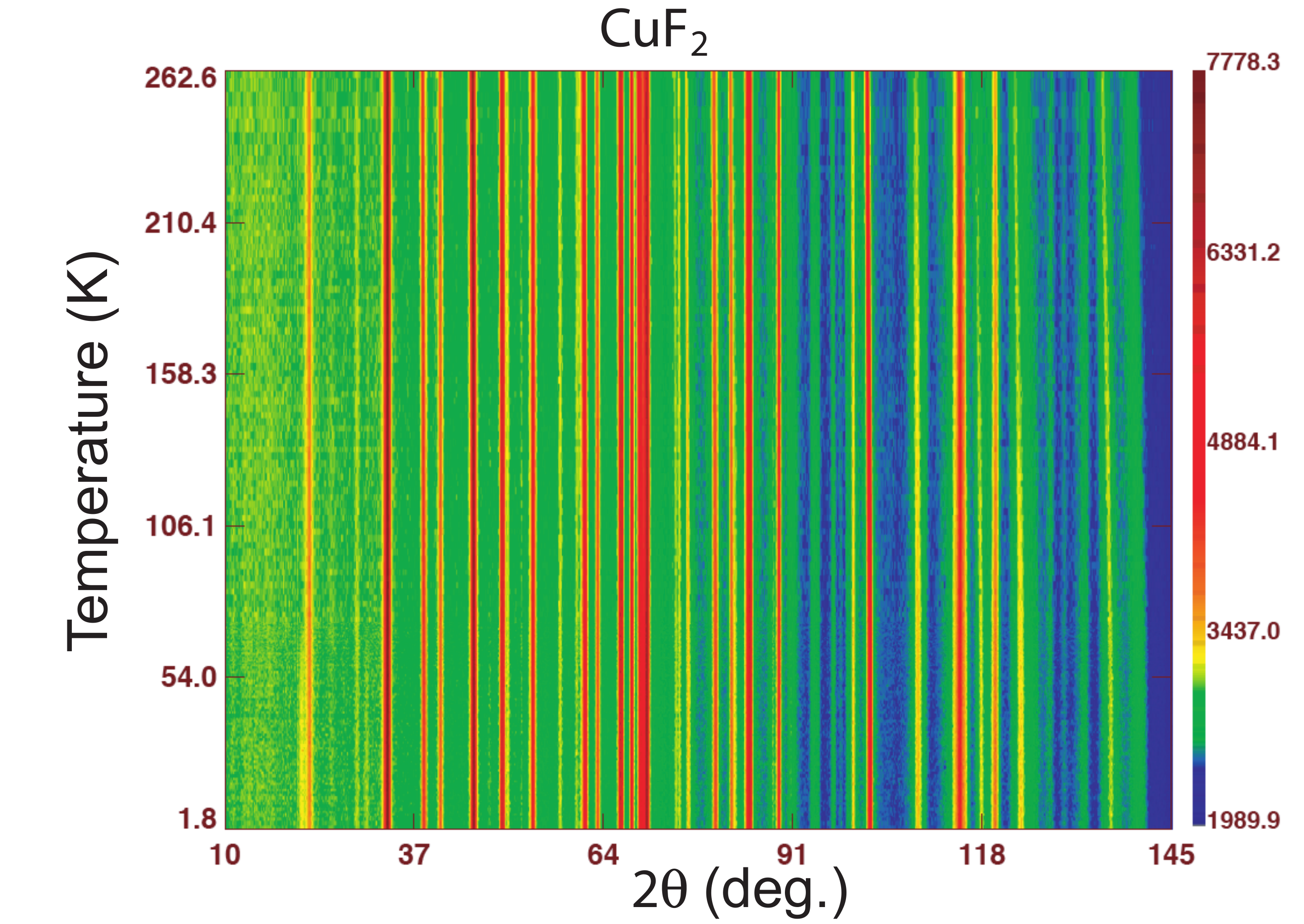}}
\caption {(Color online)Temperature variation of the diffraction diagram of CuF$_2$ which contain the impurity phase CuF$_2$.2H$_2$O.  There are two magnetic peaks at about $2\theta \approx 20$ deg. whose intensites decrease continuously with increasing temperature and become zero at $T_N \approx 70$ K . The magnetic peaks are very weak in intensity compared to those of CrF$_2$.} 
\label{lampCu}
\end{figure}

\begin{figure}
\resizebox{0.5\textwidth}{!}{\includegraphics{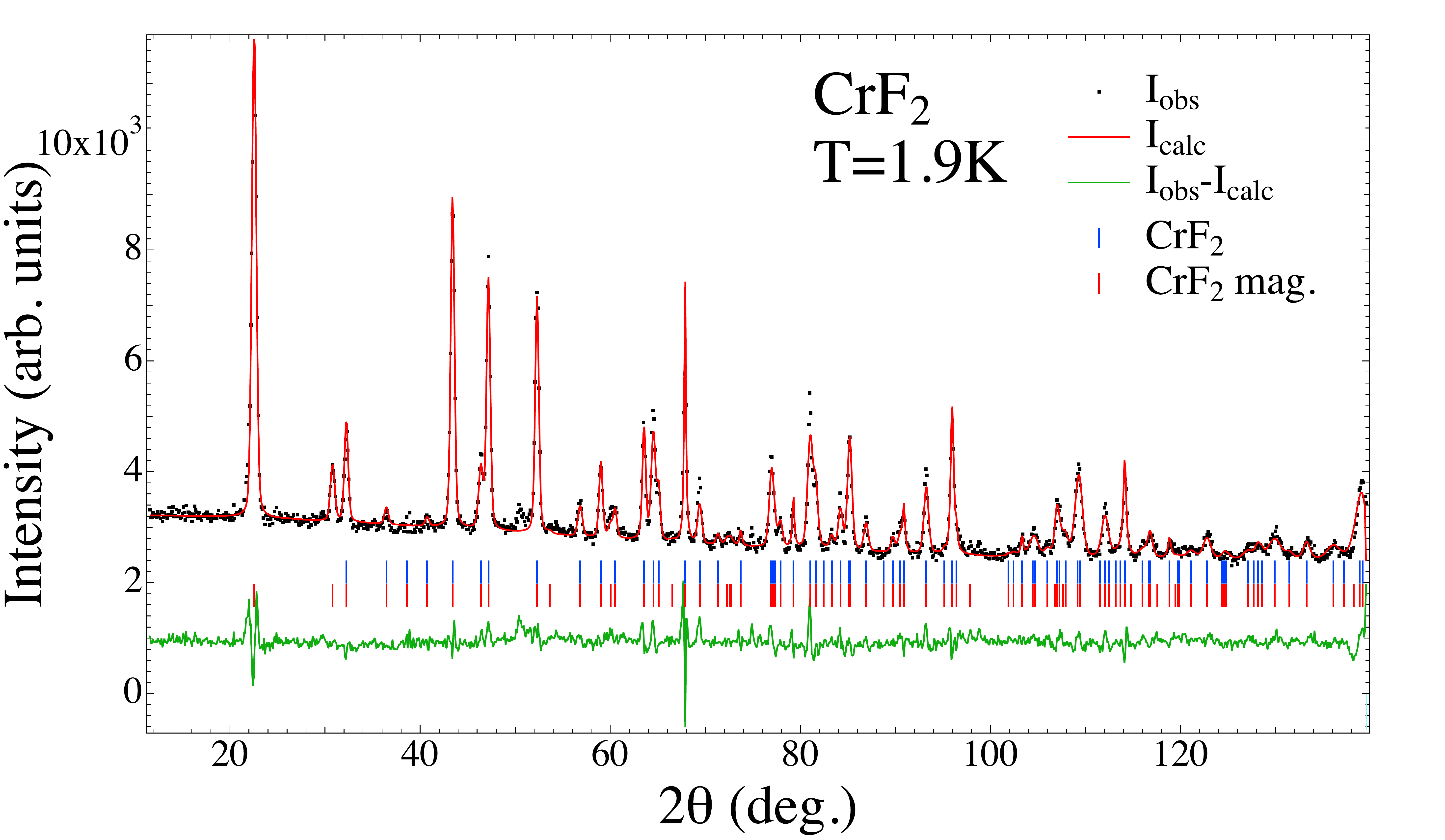}}
\caption {(Color online) The diffraction diagram of CrF$_2$ at T = 1.9 K along with the result of the Rietveld refinement of the monoclinic  $P2_1/n$ crystal structure and the  antiferromagnetic structure described in the text.} 
\label{crdiffraction}
\end{figure}

\begin{figure}
\resizebox{0.5\textwidth}{!}{\includegraphics{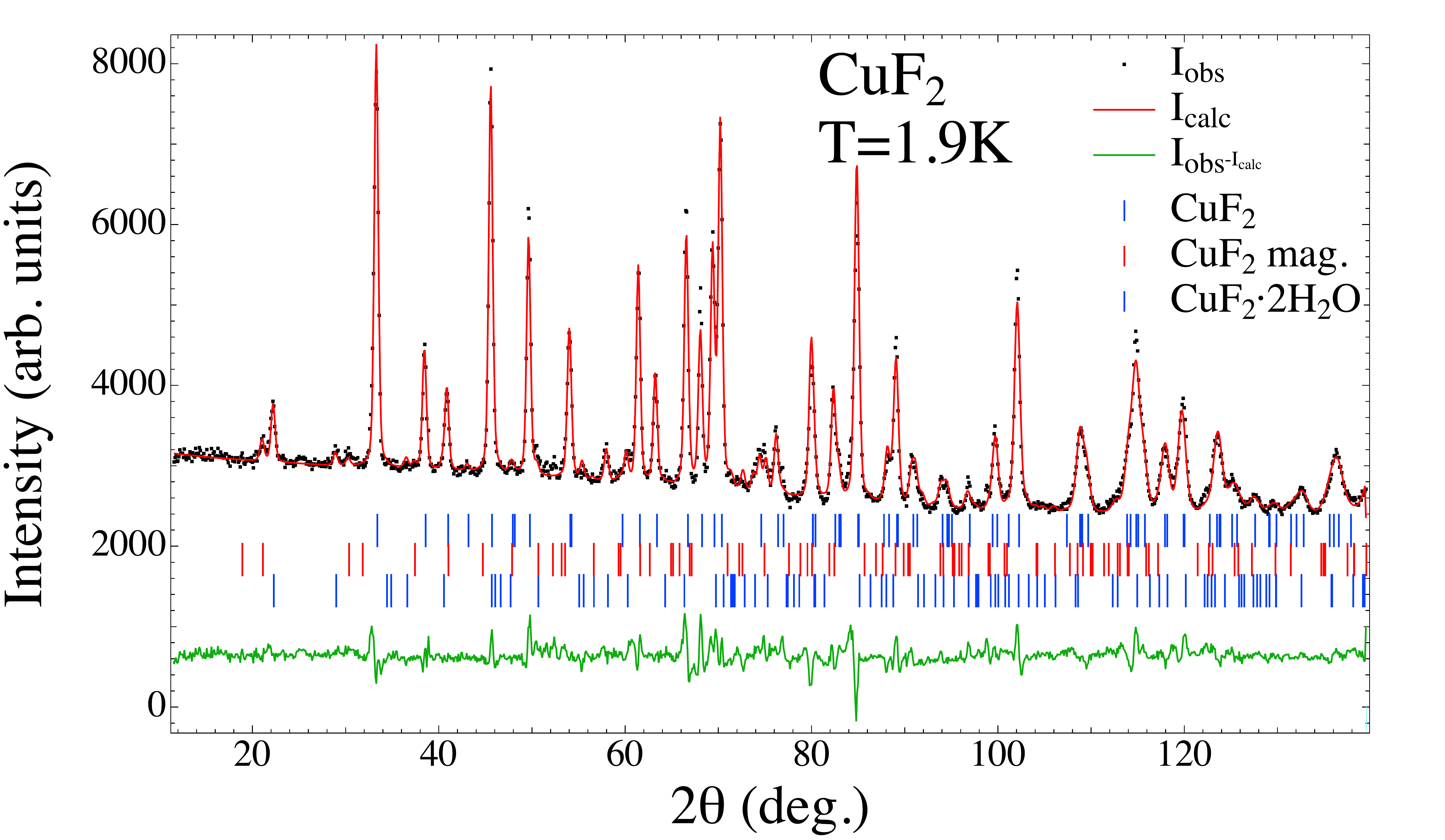}}
\caption {(Color online) The diffraction diagram of CuF$_2$ at T = 1.9 K along with the result of the Rietveld refinement of the monoclinic $P2_1/n$  crystal structure and the  antiferromagnetic structure described in the text. The impurity phase CuF$_2$.2H$_2$O has also been refined.} 
\label{cudiffraction}
\end{figure}

The transition metal fluorides constitute a class of materials that has been often investigated for checking and for the improvement of magnetic models \cite{tressaud82}. The transition metal difluorides MF$_2$ (M = V,Cr,Mn,F,Co,Ni,Cu) were investigated quite early and very intensively. Among these difluorides CrF$_2$ and CuF$_2$ are different in that they contain Jahn-Teller distorted MO$_6$ octahedra and they crystallize in the monoclinic distorted \cite{jack57,billy57} (space group $P2_1/n$) rutile structure rather than the ideal tetragonal rutile structure (space group $P4_2/mnm$) adopted by the rest. Despite the distorted rutile type crystal structure, the antiferromagnetic structure \cite{cable60} of CrF$_2$ is  similar to those of other difluordes MnF$_2$, FeF$_2$ and CoF$_2$ in which the magnetic moments of the corner metal ions at $(0,0,0)$ are aligned parallel and those at $(1/2,1/2,1/2)$ are aligned anti-parallel. The propagation vector of this type magnetic structure is ${\bf k} = (0,0,0)$. The orientation in MnF$_2$, FeF$_2$ and CoF$_2$ are parallel and anti-parallel to the tetragonal $c$ axis. The magnetic moments in CrF$_2$ are however not aligned in any crystallographic direction, but are aligned along the longest Cr-F bond directions of the Jahn-Teller distorted MF$_6$ octahedra. NiF$_2$ has also a similar magnetic structure but the moments are aligned approximately perpendicular to the tetragonal c-axis. The magnetic structure of CuF$_2$ is even more complex. VF$_2$ has also a complex helimagnetic structure. The crystal structures of CrF$_2$ is schematically illustrated in Fig. \ref{structure}. The figure also shows the Jahn-Teller distortion in CrF$_2$ by indicating the different Cr-F bond distances. The distorted octahedra have two pairs of different short  Cr-F bonds ($1.98$ and $2.01$ {\AA}) in the basal plane and a pair of long Cr-F bonds ($2.43$ {\AA}) along the axial direction. The crystal structure and Jahn-Teller distortion of CuF$_2$ are very similar to that of CrF$_2$ illustrated schematically in Fig. \ref{structure}. The two pairs of short distances in CuF$_2$ are $1.92$ and $1.93$ {\AA} with a pair of long distance $2.30$ {\AA}. The Jahn-Teller distortions in CrF$_2$ and CuF$_2$ are referred to as text book examples of this effect and have indeed been discussed in some text books {\cite{huheey83,pauling60}. The magnetic structure \cite{fischer74} of CuF$_2$ is different from that of other transition metal difluorides.  Like NiF$_2$, CrF$_2$ and CuF$_2$ also exhibit weak ferromagnetism and therefore their spin directions are canted from the ideal antiferromagnetic opposite spin orientations by small  angles. The weak ferromagnetism is allowed by the symmetry of their magnetic structures. 
However the weak ferromagnetism cannot be investigated by unpolarized neutron diffraction technique of the present study. Single-crystal polarized neutron diffraction and also laboratory magnetization investigations are necessary for such study. 

We recently investigated the magnetoelastic effects in the tetragonal transition metal difluorides \cite{chatterji10a,chatterji10b} MnF$_2$, FeF$_2$, CoF$_2$ and NiF$2$ and found some interesting systematic variation of this effect across the transition metal series especially as a function of the unquenched orbital moments. Here we have investigated the magnetoelastic effects in the Jahn-Teller distorted monoclinic CrF$_2$ and CuF$_2$. This has enabled us to check the variation of these effects across the whole transition metal series except for VF$_2$, a sample more difficult to prepare.

 \section{Experimental details}
 We have done powder neutron diffraction measurements on MF$_2$ (M=Cr,Cu) using the high-intensity two-axis powder diffractometer \cite{hansen08} D20 at the Institut-Laue-Langevin, Grenoble. The $(115)$ reflection from a Ge monochromator at a high take-off angle of $118^{\circ}$ gave a neutron wavelength of 1.868 {\AA}.  Approximately 3-5 g MF$_2$  (M=Cr,Cu) powder samples were placed inside an $8$ mm diameter vanadium can, which was fixed to the sample stick of a standard $^4$He cryostat.

 \section{Temperature dependence of crystal and magnetic structures}
  Fig. \ref{lampCr} shows the temperature dependence of the diffraction diagram of CrF$_2$. The unresolved low angle magnetic $100/010$ reflections close to the Bragg angle $20^{\circ}$ appears at low temperature. Its intensity decreases continuously as a function of temperature and becomes zero at $T_N \approx 50$ K. Fig. \ref{lampCu} shows the temperature dependence of the diffraction diagrams of CuF$_2$. The diffraction diagram contains many more peaks than are expected from the CuF$_2$ crystal and magnetic structure. A quick search for the probable impurity phase immediately convinced us that the sample contained substantial amount of CuF$_2$.2H$_2$O as the impurity phase. The magnetic peaks were very weak as expected from magnetic neutron scattering from Cu$^{2+}$ ions. However weak magnetic reflections expected from the published magnetic structure were identified. The magnetic reflections decrease in intensity with increasing temperature and become zero at about $T_N \approx 70$ K.   Figure \ref{crdiffraction} shows a typical diffraction diagram of CrF$_2$ measured at $T = 1.9$ K along with the results of Rietveld refinement of the monoclinic $P2_1/n$ crystal structure and the  antiferromagnetic structure described in the previous section. However the magnetic moment direction obtained from the present refinement of the intensity data differed from that reported by Cable et al. \cite{cable60}. Fig. \ref{cudiffraction} shows the results of the refinements of the crystal and magnetic structures of CuF$_2$  at $T = 1.9$ K along with that of the impurity phase CuF$_2$.2H$_2$O. Note that the weak ferromagnetism has not been taken into account in the refinement of the magnetic structures of CrF$_2$ and also CuF$_2$.

\begin{figure}
\resizebox{0.42\textwidth}{!}{\includegraphics{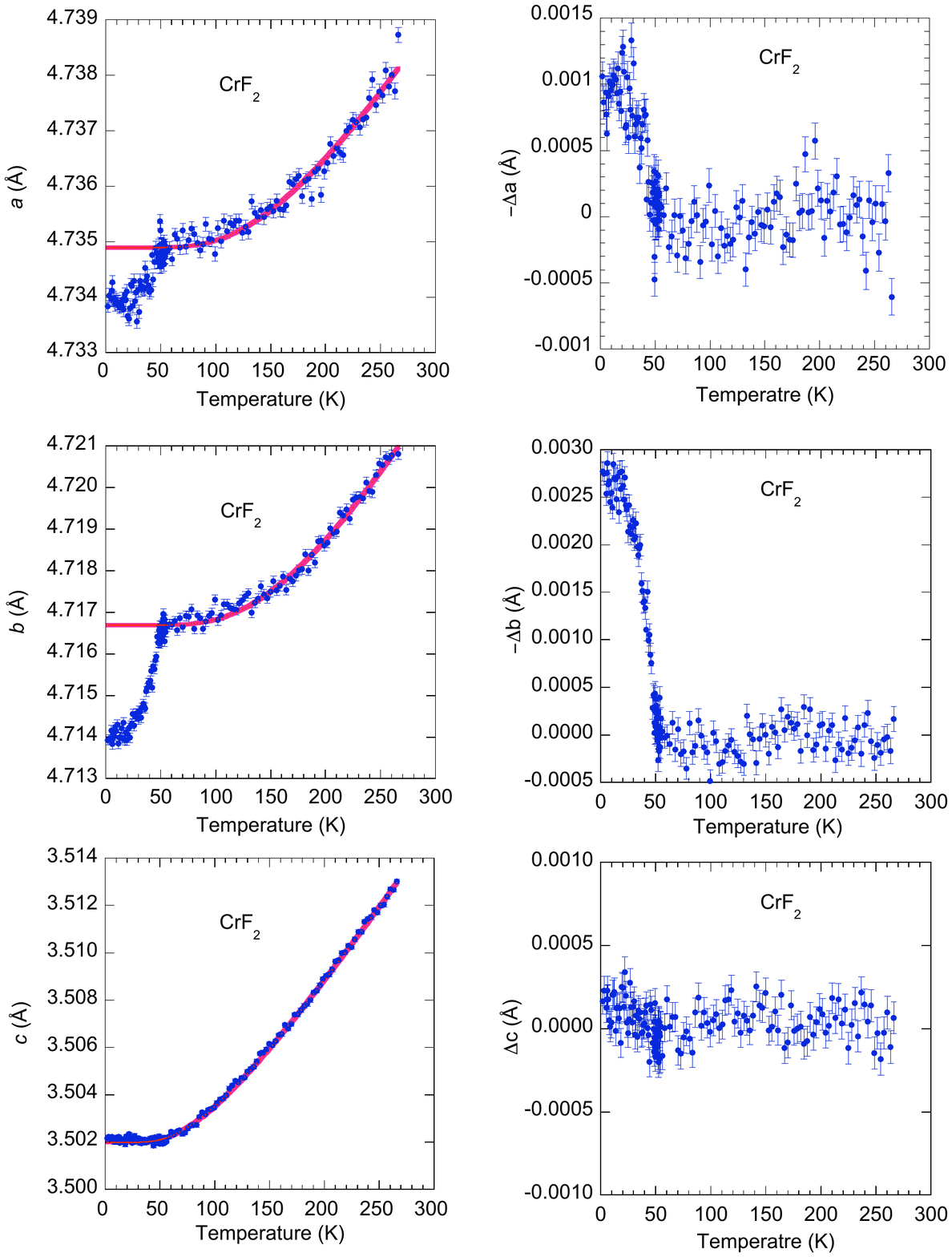}}
\resizebox{0.42\textwidth}{!}{\includegraphics{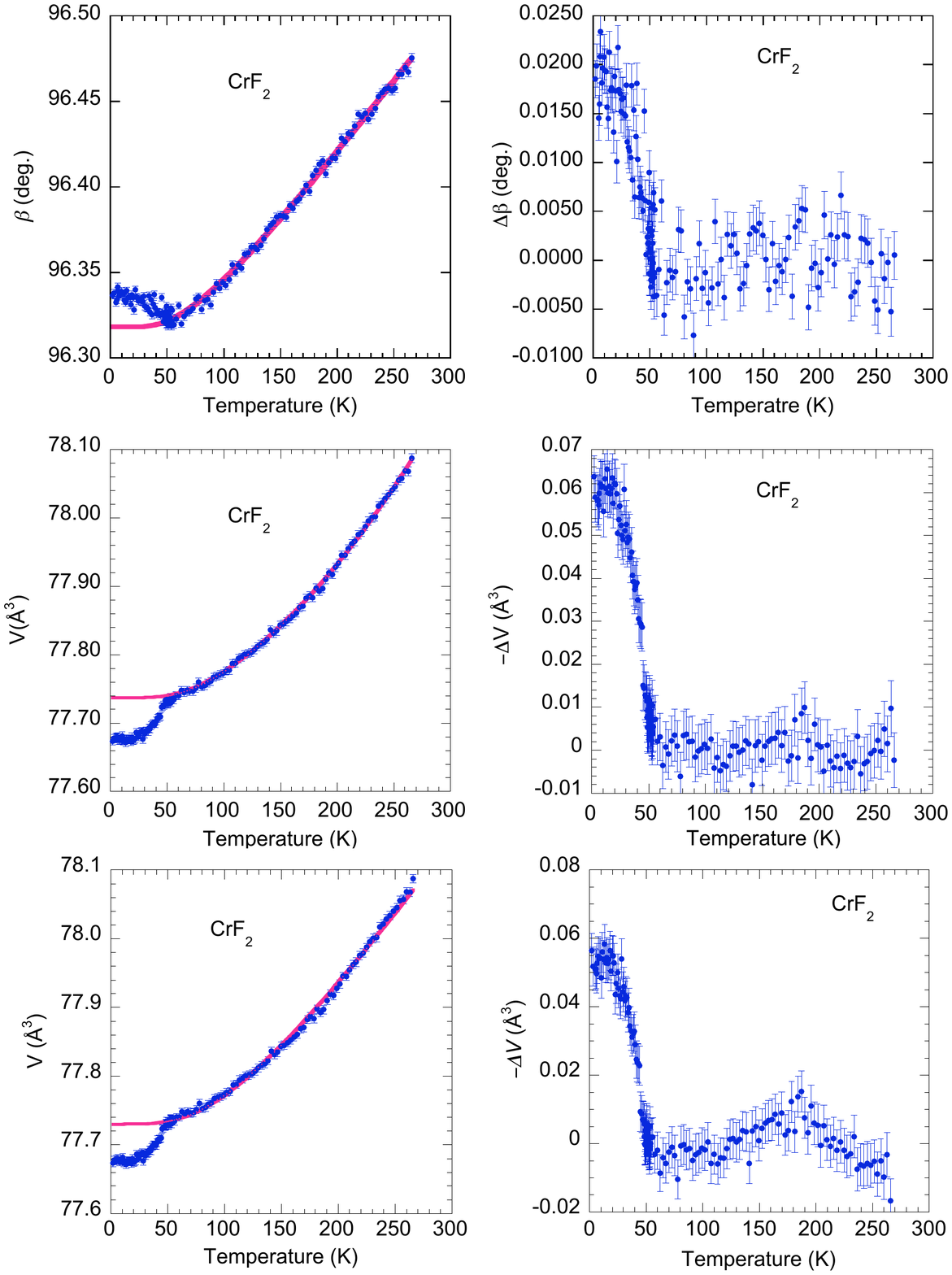}}

\caption {(Color online) Temperature variation of the lattice parameters  $a$, $b$, $c$, $\beta$ and the unit cell volume $V$  of CrF$_2$ plotted on the left side. The red curves in these figures represent the lattice parameter and the unit cell volume obtained by fitting the high temperature data in the paramagnetic state to the Einstein equation (\ref{einstein} and \ref{einstein1})  and by extrapolating to the lower temperatures in the ordered state as explained in the text and should give the corresponding values in the absence of the magnetic transition. The excess lattice parameters $\Delta a$, $\Delta b$, $\Delta c$, $\Delta\beta$ and the excess unit 
cell volume $\Delta V$ due to magnetostriction have been plotted on the right side. In the bottom panel we have shown the fit with the Debye model (equations \ref{gruen} and \ref{debye}). The resultant $\Delta V$ plotted on the right bottom panel shows that the fit is not as good as that with the Einstein model.}
\label{crflattice}
\end{figure}

\begin{figure}
\resizebox{0.42\textwidth}{!}{\includegraphics{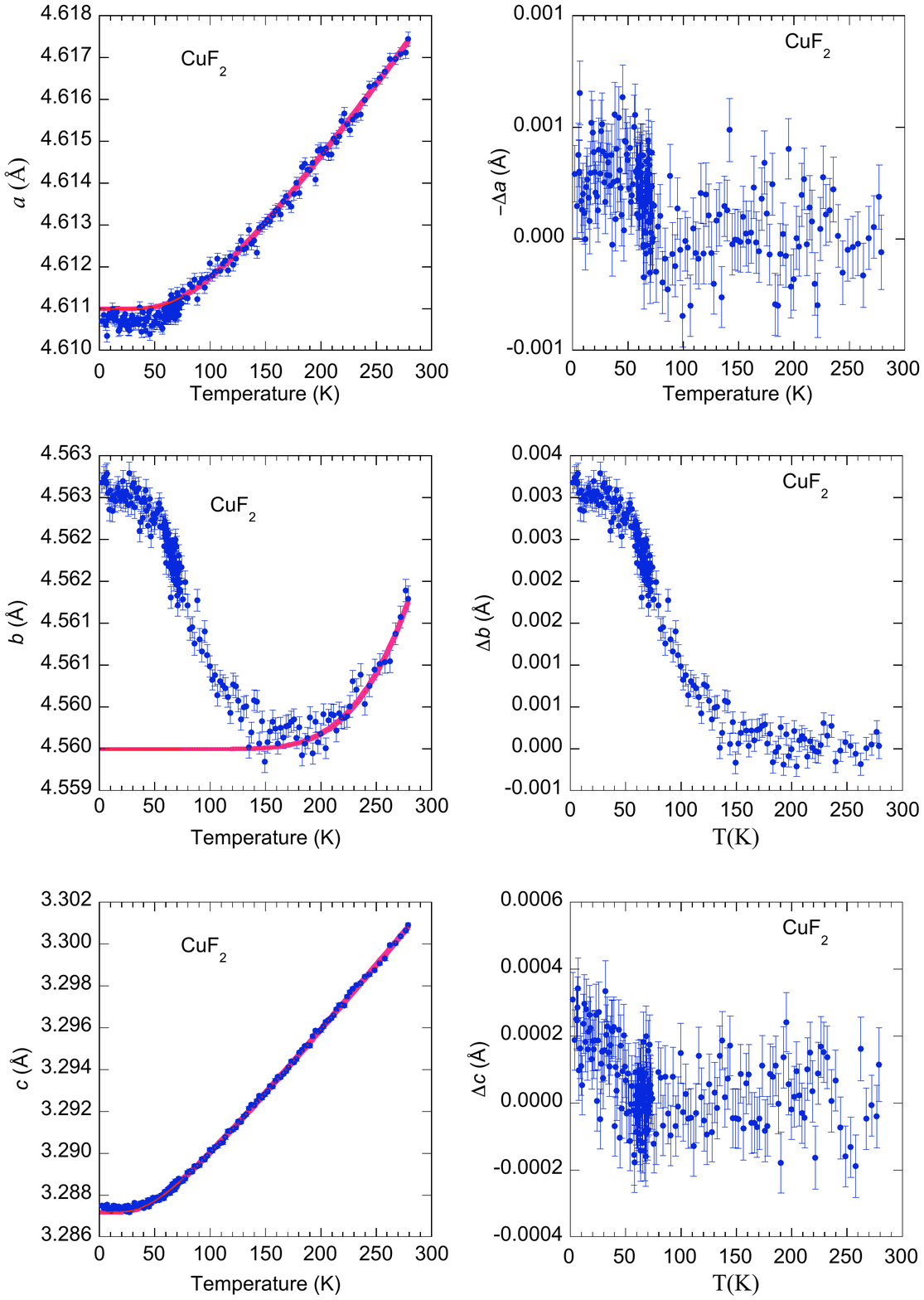}}
\resizebox{0.42\textwidth}{!}{\includegraphics{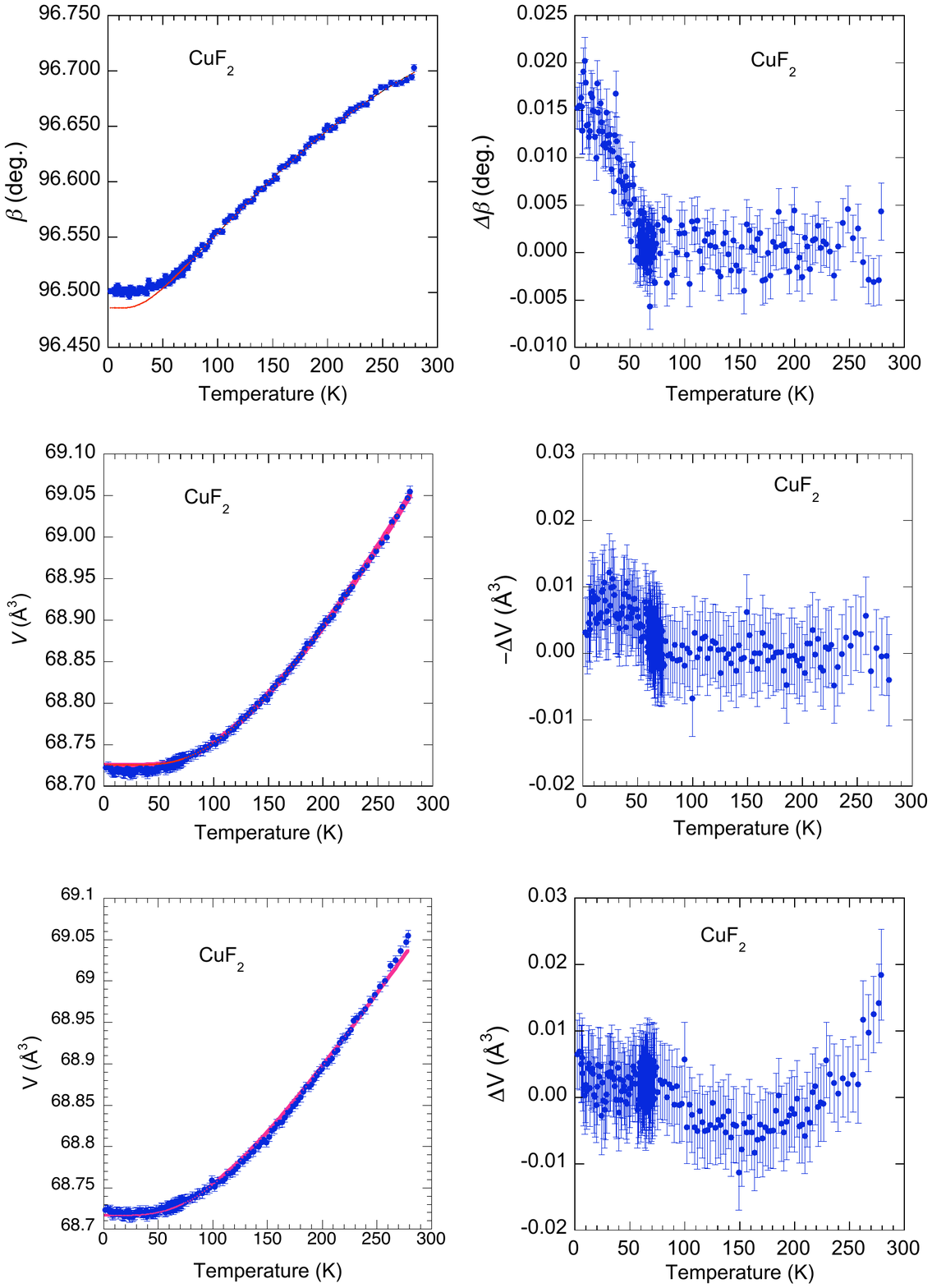}}

\caption {(Color online) Temperature variation of the lattice parameters  $a$, $b$, $c$, $\beta$ and the unit
cell volume $V$  of CuF$_2$ plotted on the left panels.  The red curves in these figures represent the lattice parameter and the unit cell volume obtained by fitting the high temperature data in the paramagnetic state to the Einstein equation (\ref{einstein} and \ref{einstein1}) and by extrapolating to the lower temperatures in the ordered state as explained in the text and should give the corresponding values in the absence of the magnetic transition. The excess lattice parameters $\Delta a$, $\Delta b$, $\Delta c$, $\Delta\beta$ and the excess unit 
cell volume $\Delta V$ due to magnetostriction have been plotted on the right panels.In the bottom panel we show the fit with the Debye model(equations \ref{gruen} and \ref{debye}). The resultant $\Delta V$ plotted on the right bottom panel shows that the fit is not as good as that with the Einstein model.}

\label{cuflattice}
\end{figure}

\
\begin{figure}
\resizebox{0.5\textwidth}{!}{\includegraphics{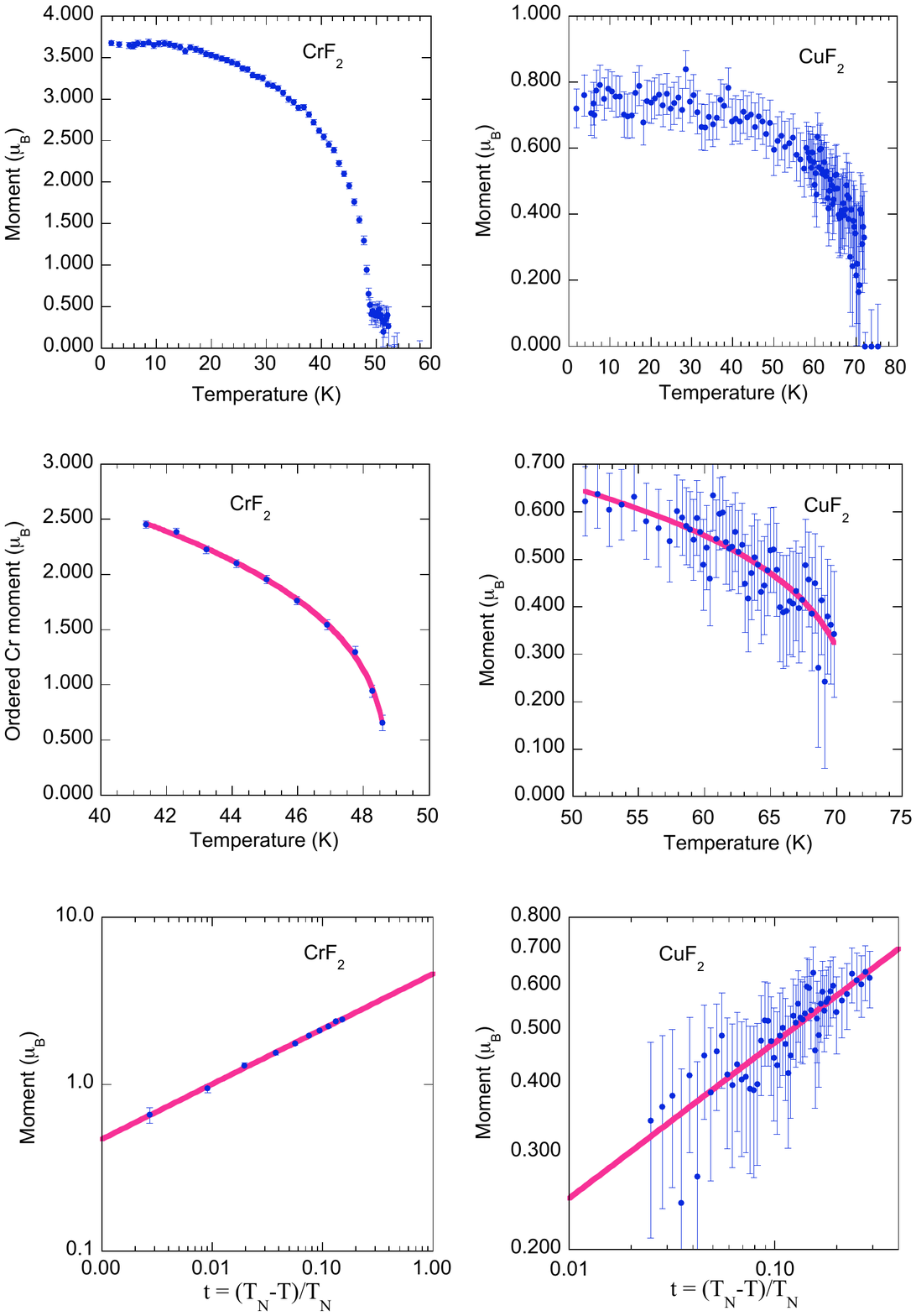}}
\caption {(Color online) (Left) Temperature variation of the ordered magnetic moment of Cr ion in CrF$_2$ in the full temperature range (top), in the critical region close to $T_N$ with a power law fit shown by the continuous curve (middle) and its log-log plot with reduced temperature (bottom). (Right) Similar plots for CuF2}
\label{magneticmoment}
\end{figure}

\begin{figure}
\resizebox{0.5\textwidth}{!}{\includegraphics{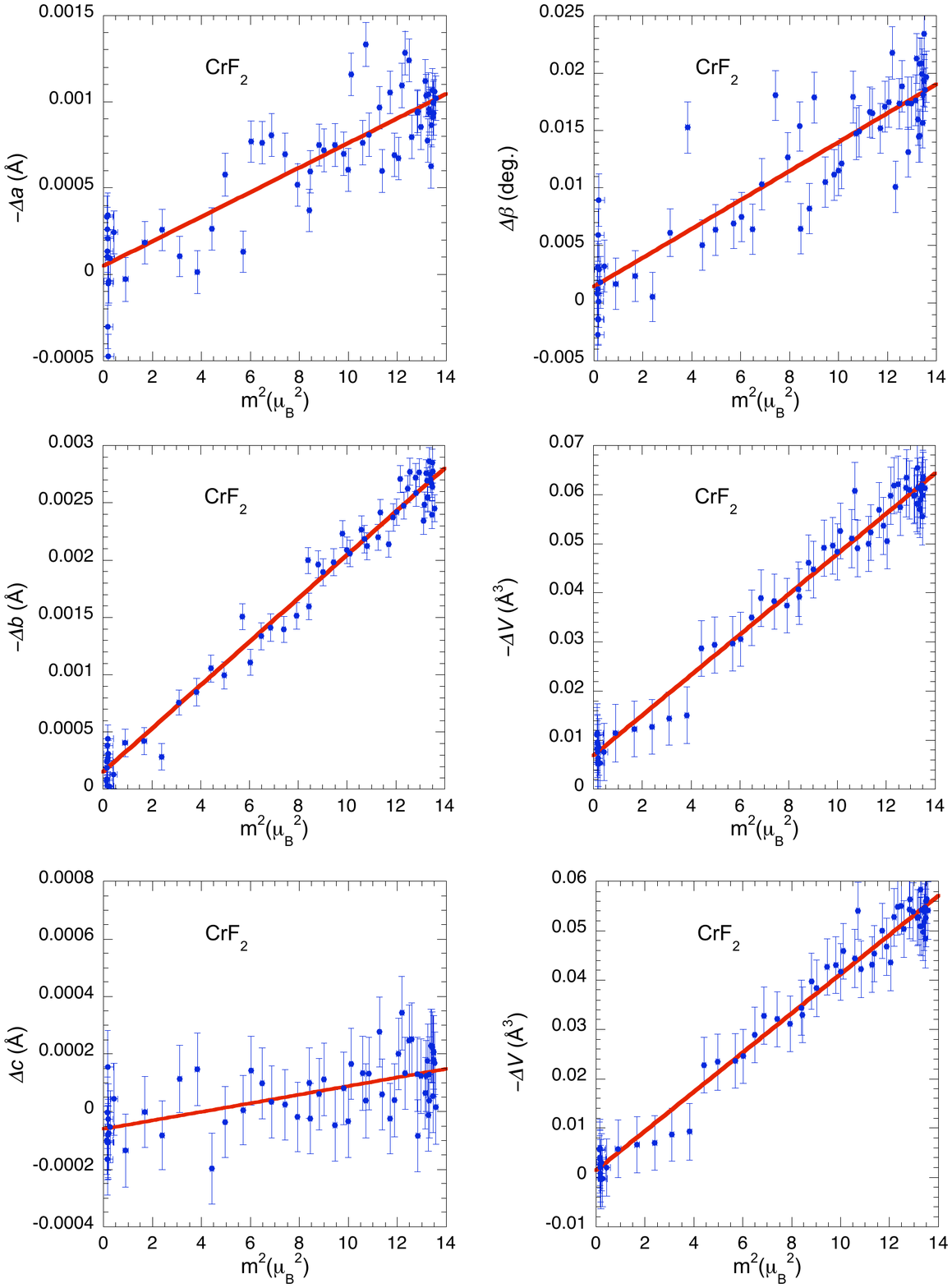}}
\caption {(Color online) Plot of the lattice strains  $\Delta a$, $\Delta b$, $\Delta c$ $\Delta \beta$ and $\Delta V$ against square of the ordered magnetic moment of Cr ion in CrF$_2$.}
\label{crcoupling}
\end{figure}

\begin{figure}
\resizebox{0.5\textwidth}{!}{\includegraphics{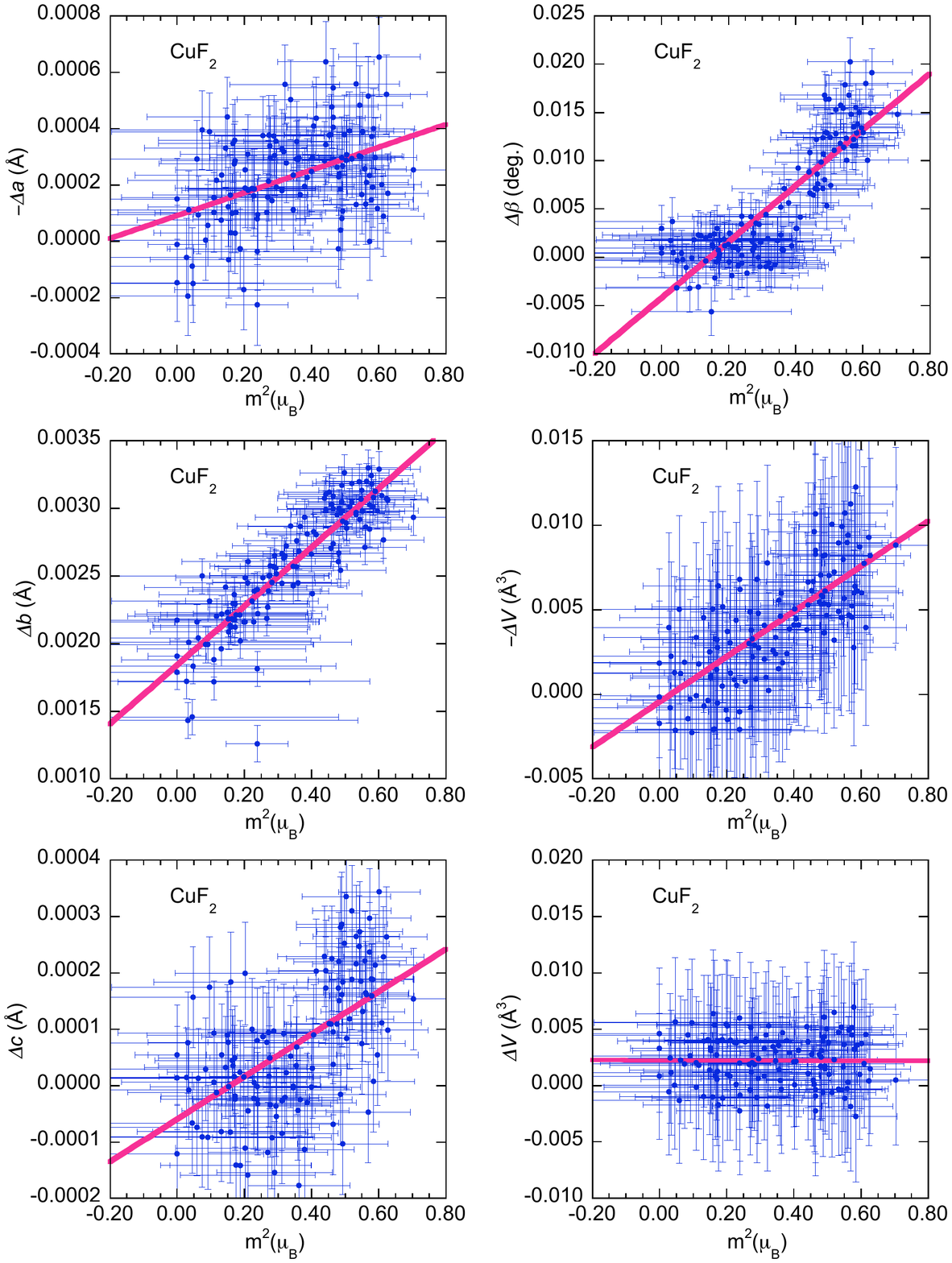}}
\caption {(Color online) Plot of the lattice strains  $\Delta a$, $\Delta b$, $\Delta c$ $\Delta \beta$ and $\Delta V$ against square of the ordered magnetic moment of Cu ion in CuF$_2$. }
\label{cucoupling}
\end{figure}

\begin{figure}
\resizebox{0.5\textwidth}{!}{\includegraphics{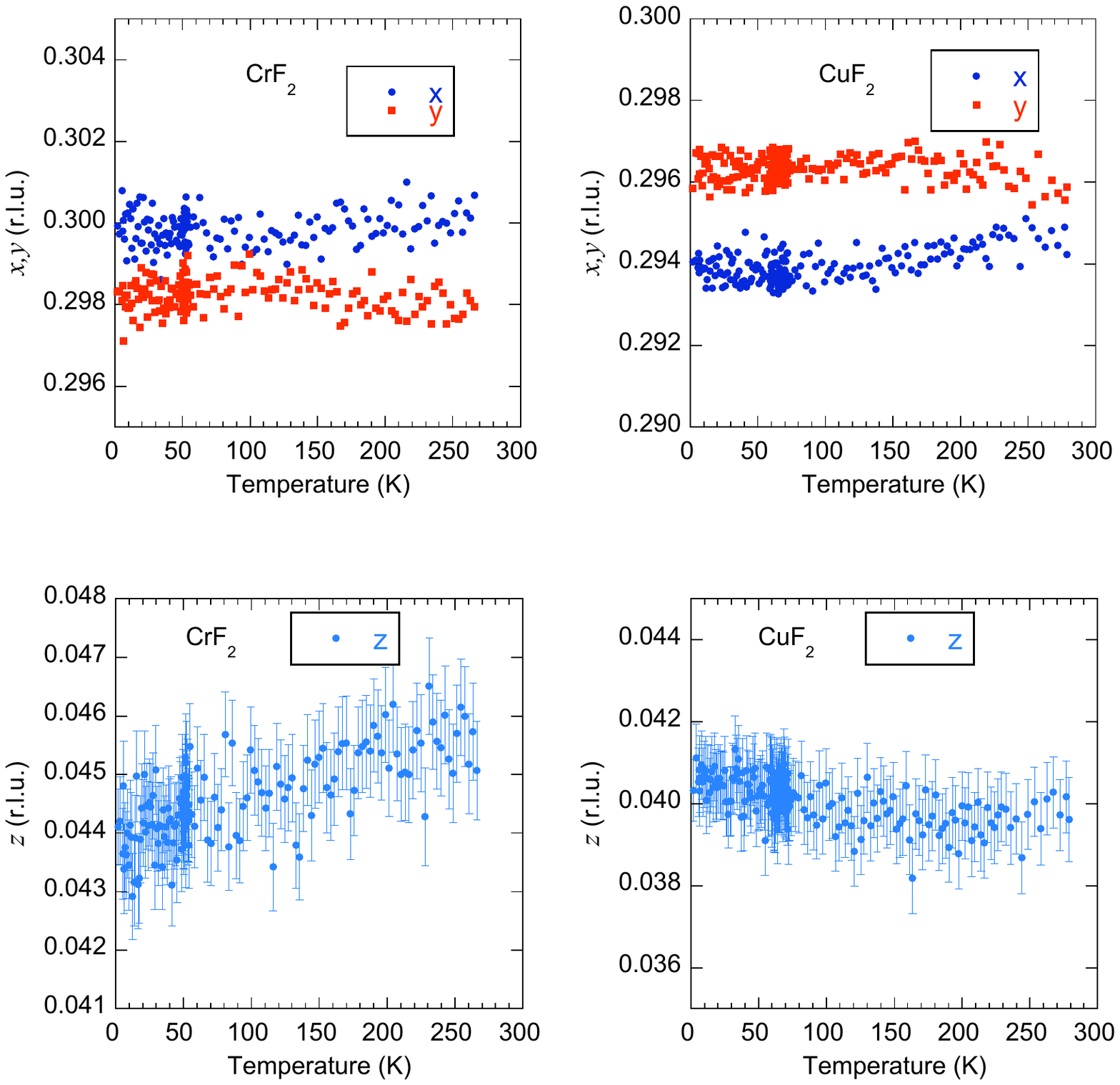}}
\caption {(Color online) (Left panel) Temperature variation of the positional parameters $x, y$ and $z$ of F atom in CrF$_2$. (Right panel)Temperature variation of the positional parameters $x, y$ and $z$ of F atom in CuF$_2$.  }
\label{crcufparameter}
\end{figure}

\begin{figure}
\resizebox{0.5\textwidth}{!}{\includegraphics{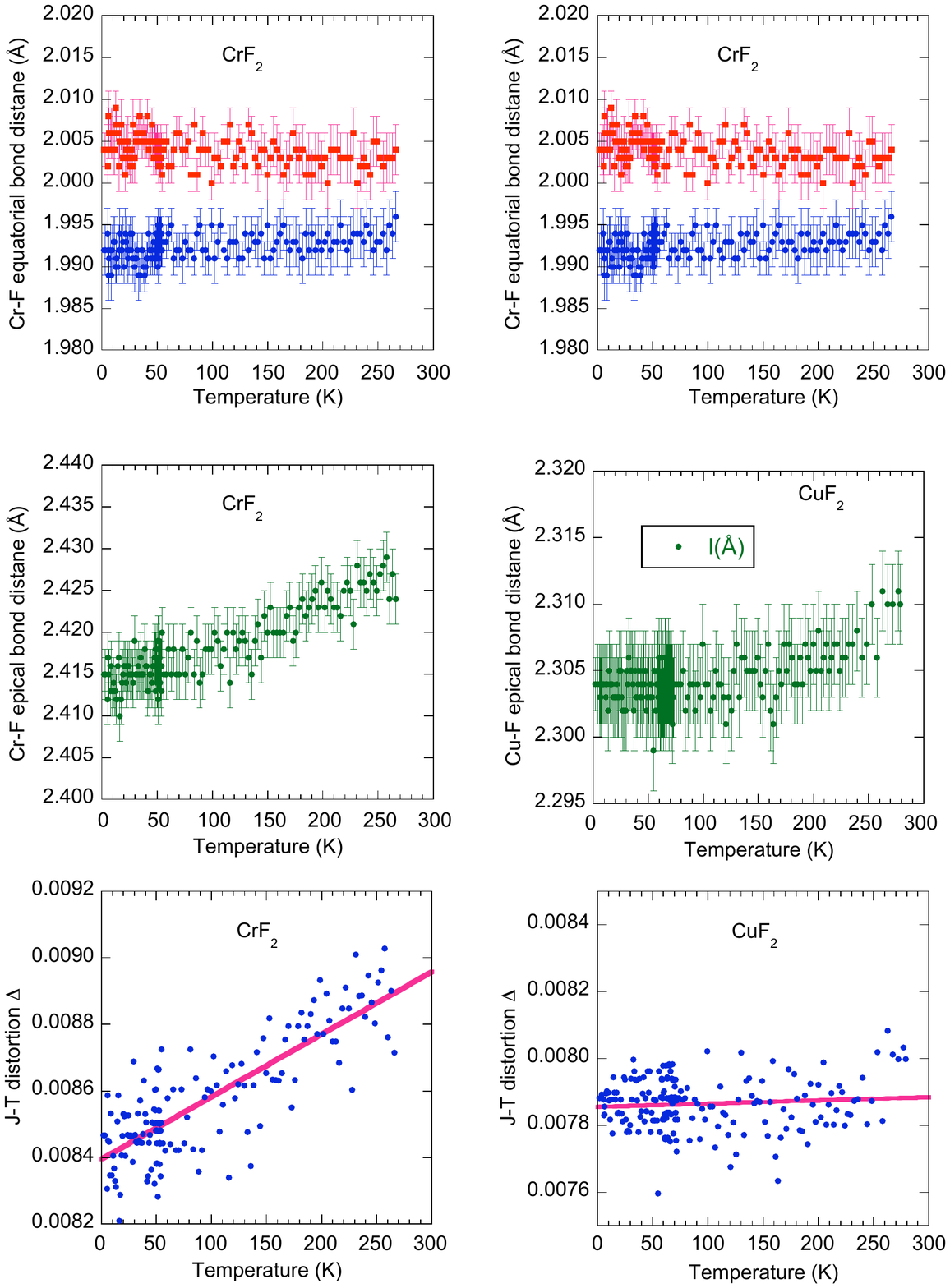}}
\caption {(Color online) (Left panel)Temperature variation of the equatorial and epical Cr-F bond distances  and Jahn-Teller distortion $\Delta$ of CrF$_2$. (Right panel)Temperature variation of the equatorial and epical Cr-F bond distances  and Jahn-Teller distortion $\Delta$ of CuF$_2$.}
\label{crcufbond}
\end{figure}

\section{Thermal expansion of lattice in absence of magnetism}
In order to study the spontaneous magnetostriction it is necessary to determine the temperature variation of the lattice parameters and the unit cell volume in the absence of magnetism. One method of  determining the background temperature variation of the lattice parameter and the unit cell volume is to extrapolate the paramagnetic high temperature data to low temperature by fitting with a polynomial function. This method works approximately in some cases but in general involves some uncertainty. Alternatively one can use the Gr\"uneisen approximation for the zero pressure equation of state, in which the effects of thermal expansion are considered to be equivalent to elastic strain \cite{wallace72}. Thus, the temperature dependence of the volume can
be described by 
\begin{equation}
V(T) =\gamma U(T)/B_0 +V_0, 
\label{gruen}
\end{equation}
where
$\gamma$ is a Gr\"{u}neisen parameter, \emph{B}$_0$ is the
bulk modulus and \emph{V}$_0$ is the volume at \emph{T} = 0 K.
By adopting the Debye approximation, the internal energy
\emph{U}(\emph{T}) is given by:
\begin{equation}
\label{debye}
{\emph{U}(T)}= 9Nk_BT
 \left(
   \frac{T}{\theta_D}
 \right)^3\int_0^{\theta_D/T}\frac{x^3}{e^x-1}\mathrm{d}x
\end{equation}
where \emph{N} is the number of atoms in the unit cell, \emph{k}$_B$
is the Boltzmann's constant and $\theta$$_D$ is the Debye
temperature.  By fitting the high temperature data in the paramagnetic state we can get the physical parameters $\theta_D$, $V_0$ and $\gamma/B_0$, which can then be used to calculate the low temperature background curve. This however, is not satisfactory because the thermal expansions of a paramagnetic and a nonmagnetic solid are generally not identical. We have used in the present study the Gr\"uneisen approximation given by the equation (\ref{gruen}). This equation is of course valid for the volume in the isotropic case. However in the present experiment we actually measure the lattice parameters $a$, $b$, $c$ and the monoclinic angle $\beta$ and the unit cell volume is only calculated from them from $V = abc\sin \beta$. The volume data  can be fitted directly by using  equation (\ref{gruen}, \ref{debye}).  However we also wish to fit the individual lattice parameters. We found out that Einstein's model serves as a better method for fitting the high temperature data to a function appropriate for the thermal expansion of a non-magnetic solid. We start from the Gr\"uneisen relation 
\begin{equation}
\gamma=\frac{\alpha_VVB_T}{C_V},
\end{equation}
where $\alpha_V$ is the volume thermal expansion and $C_V$ is specific heat at constant volume and $V$ is the molar volume. From the assumption of temperature independence of $\gamma$ and $B_T$ we can express the volume as 
\begin{equation}
\label{volume}
V(T)=V_0+\frac{\gamma}{B_T}\int_0^T C_V.
\end{equation}
Now using the Einstein model for the specific heat in which all atoms in the solid are assumed to vibrate with the same characteristic angular frequency $\omega_E$ where $\theta_E=\hbar\omega_E/k_B$ we have
\begin{equation}
C_V=\frac{3R(\theta_E/T)^2\exp{(\theta_E/T)}}{(\exp{(\theta_E/T)}-1)}
\end{equation}
Substituting $C_V$ in equation \ref{volume} and integrating
\begin{equation}
\label{volume1}
V(T)=V_0+\frac{E}{(\exp{(\theta_E/T)}-1)}
\label{einstein}
\end{equation}
where the constant $E = 3R\gamma\theta_E/B_T$. A similar expression can be used to describe the temperature dependence of lattice parameters,
\begin{equation}
x(T)=x_0+\frac{E}{(\exp{(\theta_E/T)}-1)}
\label{einstein1}
\end{equation}
where $x_0$ is the value of the lattice parameter $x$ ($a,b,c,\beta$ in our case) at $T = 0$ K. In order to get a good fit sometime we need to allow the parameter $E$ to vary as a function of temperature as a polynomial in $T$.
Although the the present Einstein model is not as rigorous as the Debye model, it is surprisingly much better as a fit function to the data as will be evident in the next section.

\section{Magnetostriction in ${\bf CrF_2}$ and ${\bf CuF_2}$}
Figure \ref{crflattice} (left) shows the temperature variation of the lattice parameters  $a$, $b$, $c$, the monoclinic angle $\beta$ and the unit cell volume $V$  of CrF$_2$ and the corresponding fit of the high temperature data above $T_N$ using Einstein model within Gr\"uneisen approximation extrapolated to the lowest temperature and are shown by the continuous curves. The lattice parameter $a$ of CrF$_2$ decreases continuously with decreasing temperature following the continuous curve and then shows strong magnetoelastic effect below $T_N$. The magnetoelastic effect obtained by subtracting the nonmagnetic background is shown in the corresponding right panel. The effect is negative i.e.  the lattice parameter decreases at the magnetic phase transition from the value expected for a nonmagnetic solid given by the continuous curve. The lattice parameter $b$ behaves in a similar way whereas the lattice parameter $c$ does not show any appreciable magnetoelastic effects. The monoclinic angle $\beta$ shows opposite positive effect. The unit cell volume shows negative magnetovolume effect. The bottom panel shows the corresponding fit of the temperature variation of the volume by using Debye model. It is seen from the corresponding plot of $\Delta V$ in the right panel that this fit of the data with the Debye model is much less satisfactory than that using Einstein equation. Figure \ref{cuflattice} shows similar plots of the unit cell parameters and volume of CuF$_2$. The lattice parameter $a$ of CuF$_2$ shows small negative effect whereas the lattice parameter $b$ shows strong positive effect. The lattice parameter $c$ shows very small negative magnetoelastic effect. The monoclinic angle $\beta$ of CuF$_2$ shows positive magnetoelastic effect. The resultant magnetovolume effect of CuF$_2$ is very small and negative. The fit with the Debye model is also shown in bottom panel and effect is again very small but positive. The right panel of Fig. \ref{cuflattice} shows $\Delta a$, $\Delta b$, $\Delta c$, $\Delta \beta$  for fits with Einstein models and $\Delta V$ for both Einstein and Debye models.

\section{Antiferromagnetic phase transitions in ${\bf CrF_2}$ and ${\bf CuF_2}$}
We investigated the antiferromagnetic phase transition in CrF$_2$ and CuF$_2$. Figure \ref{magneticmoment} (left) shows the temperature variation of the ordered magnetic moment of Cr ion in CrF$_2$ in the full temperature range on the top panel obtained from the refinement of magnetic intensities by using the magnetic structure model of Cable et al. \cite{cable60}. The continuous decrease of the magnetic moment with increasing temperature shows the second order nature of the antiferromagnetic phase transition in CrF$_2$. The data in the critical range close to $T_N$ is shown in the middle left panel along with the least squares fit with a power law exponent
\begin{equation}
m(T)=m_0\left(\frac{T_N-T}{T_N}\right)^\beta.
\end{equation}
shown by the continuous curve. The critical exponent was determined to be $\beta = 0.34 \pm 0.09$ which agrees closely with the expected three-dimensional Heisenberg value\cite{collins89} $0.367$ within experimental accuracy. The N\'eel temperature $T_N = 48.7 \pm 0.1$ K determined from the fit agrees with that reported in the literature \cite{cable60}. The log-log plot of the magnetic moment with the reduced temperature $t= \frac{T_N-T}{T_N}$ is shown in the bottom left panel. Fig. \ref{magneticmoment} (right panel) shows similar results for the antiferromagnetic phase transition in CuF$_2$. The N\'eel temperature was determined to be $T_N = 72 \pm 9$ K which agrees within experimental accuracy with that reported in literature \cite{fischer74}. However due to much larger experimental errors in the refined magnetic moment values the critical exponent $\beta$ of CuF$_2$ could not be determined with any reasonable accuracy. The magnetic moments determined by the refinements of the intensity data were $3.6 \pm 0.2$ and $0.75 \pm 0.10 \mu_B$  for CrF$_2$ and CuF$_2$, respectively.

\section{Coupling of the lattice strain with the order parameter}
In order to check how the lattice strains $\Delta a$, $\Delta b$, $\Delta c$, $\Delta \beta$ and $\Delta V$  couple to the order parameter these were plotted against the square of ordered magnetic moment obtained from the refinement of the magnetic structure. Figure \ref{crcoupling} shows the corresponding plots for CrF$_2$ in the left and right panels. For $\Delta V$ there are two plots corresponding to the two different background fits from Einstein and Debye equations. The ordered magnetic moment is the order parameter. The plots show that all these lattice strains  couple linearly with the square of the order parameter for CrF$_2$. The linearity is quite obvious for $\Delta b$ and $\Delta V$ but it is less so for $\Delta a$, $\Delta c$ and $\Delta \beta$ due to the small magnetovolume effect and also large error bars. Fig. \ref{cucoupling} shows similar plots for CuF$_2$ on the left and right panels but due to the very large errors of the data nothing much can be concluded from the plots except for that for $\Delta b$ for which linearity is definitely valid. The quadratic relationship between the lattice strain and the order parameter is expected for a single sublattice ferromagnet \cite{clark80}.  Chatterji et al. \cite{chatterji10a,chatterji10b} showed recently that the quadratic coupling between the lattice strain and the order parameter is also valid for CoF$_2$, MnF$_2$, FeF$_2$ and NiF$_2$. Thus the coupling of the lattice stain is clearly quadratic to the order parameter in the transition metal difluoride series. This is consistent with the simple argument given in References \cite{andreev95,clark80}. 

\section{Jahn-Teller distortion in ${\bf CrF_2}$ and ${\bf CuF_2}$}
As we mentioned before the CrF$_2$ and CuF$_2$ stand out from the rest of transition metal difluorides in that they contain Jahn-Teller ions Cr$^{2+}$ and Cu$^{2+}$ that cause the distortion of the CrF$_6$ and CuF$_6$ octahedra containing four short and two long Cr-F and Cu-F bond distances. In order to check whether the antiferromagnetic phase transition modifies the bond distances at low temperatures we determined the lattice positional parameters as functions of temperature. Figure \ref{crcufparameter} gives the temperature variation of the positional parameters $x$, $y$ and $z$ of the F atom of CrF$_2$ in the left panel and those for CuF$_2$ in the right panel. We do not observe any drastic change in the positional parameters close to the antiferromagnetic phase transitions for both CrF$_2$ and CuF$_2$. Fig. \ref{crcufbond} shows the temperature variation of short equatorial and long apical bond distances in the CrF$_6$ octahedra of CrF$_2$ in the left panel and those for CuF$_2$ in the right panel. In bottom panels are shown the Jahn-Teller distortion parameter $\Delta$ as a function of temperature. The distortion parameter $\Delta$ of a coordination polyhedron AB$_N$ with an average A-B bond distance $<d>$ and the individual A-B distances $d_n$ is defined as
\begin{equation}
\Delta=\frac{1}{N}\sum_{n=1,N}\left(\frac{d_n-<d>}{<d>}\right)^2.
\end{equation}
 We note that apart from smooth linear decrease with temperature no drastic changes in bond distances take place at the antiferromagnetic phase transition in CrF$_2$ and CuF$_2$. The epical Cr-F distance of the octahedron decrease linearly with decreasing temperature in CrF$_2$ whereas the equatorial Cr-F bond distances remain almost constant in the temperature range investigated. There is hardly any effect in these bond distances in CuF$_2$. Also the distortion parameter $\Delta$ of the octahedron decreases linearly with decreasing temperature in CrF$_2$ where $\Delta$ remains more or less constant  in CuF$_2$ in the temperature range studied. There is no magnetoelastic effect of the Jahn-Teller distortion at the transition temperature. This is what is expected due to the much higher energy scale of the Jahn-Teller effect compared to the magnetic exchange and magnetoelastic energies.
 
  \begin{table}[ht]
\caption{Magnetic and magnetoelastic properties of MF$_2$ }
\label{table1}
\begin{center}
\begin{tabular}{ccccc} \hline \hline
\emph{Compound} & $T_N$ (K)&\emph{Moment} ($\mu_B$)&$10^{4} \times\Delta V/V$  & \emph{Reference}\\ \hline
CrF$_2$ & 48.7(1)&3.6(2)&-7.72&present work\\
MnF$_2$ & 69.0(2)&5.12(9)&-10.02&[15]\\
FeF$_2$ & 79(1)&4.05(5)&-4.1&[15]\\
CoF$_2$ & 39.0(4)&2.57(2)&2.6&[14]\\
NiF$_2$ & 71.0(6)&1.99(5)&-4.5&[15]\\
CuF$_2$ & 72(9)&0.75(10) &-1.16&present work\\ \hline
\end{tabular}
\end{center}
\end{table}

\begin{figure}
\resizebox{0.5\textwidth}{!}{\includegraphics{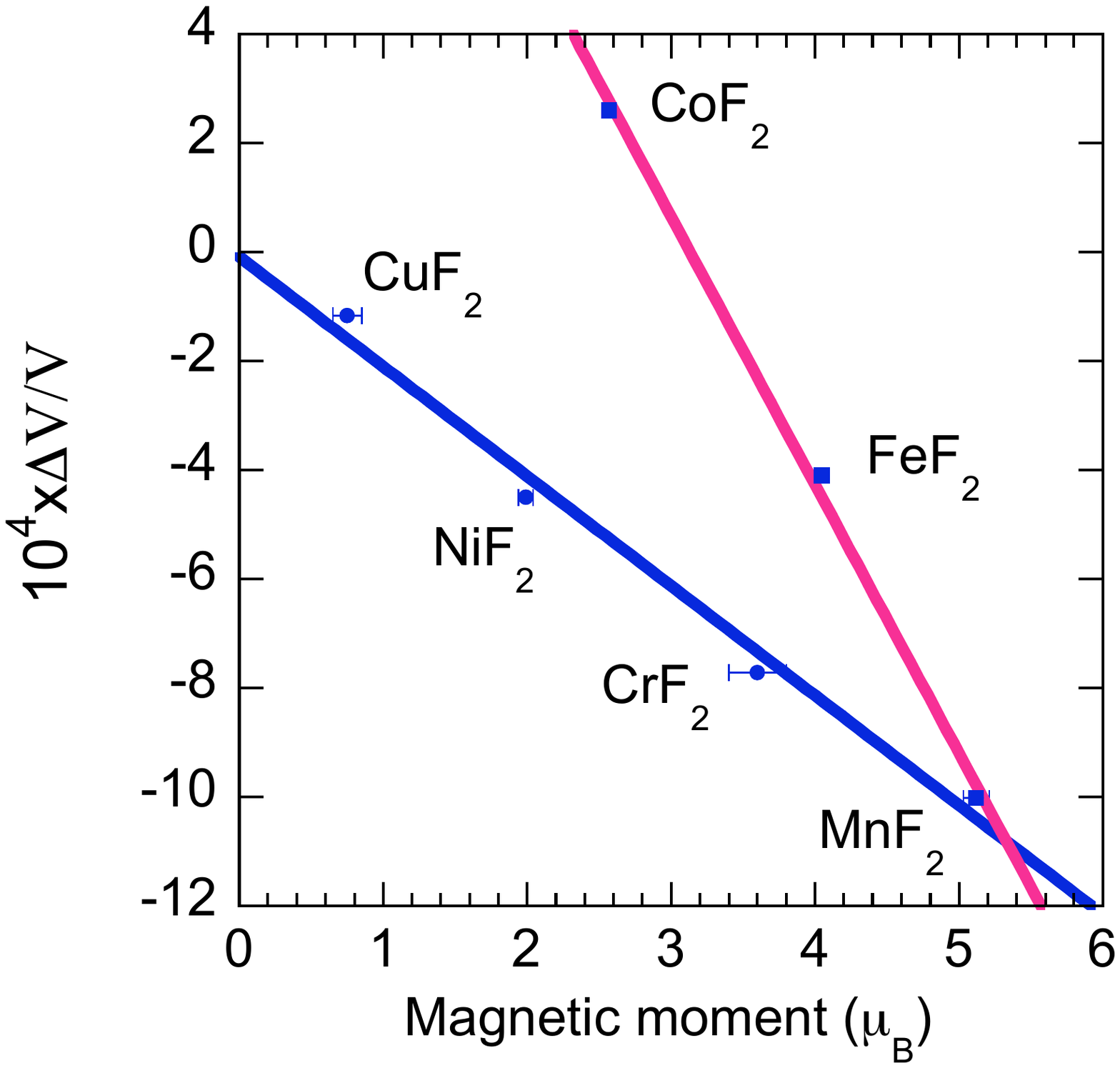}}
\caption {(Color online) Plot of the magnetovolume effect $\Delta V/V$ vs. the magnetic moment.}
\label{correlation}
\end{figure}

\section{Magnetovolume effect}
We now consider the correlation of  the magnetovolume effects at the antiferromagnetic phase transition in MF$_2$
(M = Cr, Mn, Fe, Co, Ni and Cu) with other magnetic properties. Table \ref{table1} summarises the results  obtained so far for the spontaneous volume magnetostriction of transition metal difluorides along with the values for the N\'eel temperature $T_N$ and the ground state ordered magnetic moment. The data for CoF$_2$ have been taken from Chatterji et al. \cite{chatterji10a} and the data for MnF$_2$, FeF$_2$ and NiF$_2$ have been taken from Chatterji et al. \cite{chatterji10b}. We have however refitted the high temperature volume data of CoF$_2$ to the Debye-Gr\"uneisen equation (\ref{debye} and \ref{gruen}) and extrapolated to the lower temperatures. From Table \ref{table1} we note that the magetovolume effect in MnF$_2$ is relatively large and negative  whereas that in CoF$_2$ is smaller but positive  whereas those for CrF$_2$, FeF$_2$, NiF$_2$ and CuF$_2$ are negative  with intermediate values. One may expect that the magnetovolume effect in the transition metal difluoride series  be related with the magnetic moment values or the exchange interaction that is proportional to the N\'eel temperature but from looking at Table \ref{table1} we see that the magnetoelastic effect has no or little correlation with the N\'eel temperature or the exchange interaction. However there may be some correlation between the magnetovolume effect and the ordered magnetic moment. Fig. \ref{correlation} shows the plot the magnetovolume effect $\Delta V/V$ vs. the magnetic moment. We could draw two separate straight lines through the data points as shown in Fig. \ref{correlation}. The magnitude of the magnetoelastic effect seems to increase with the magnetic moment value for each of these two groups albeit with different slopes. The data for MnF$_2$ falls on both these two straight line. The implication of the existence of two straight lines with two different slopes, i.e two different rates of increase of the magnetovolume effect with magnetic moment is not quite clear. One is tempted to ascribe these two different classes of compounds with pure spin moment and spin plus orbital moment. However this is just a conjecture only and needs to be verified. This however is  consistent with our earlier observation\cite{chatterji10b} about the dependence of magnetovolume effect on the orbital momentum. Since orbital moment values are not known with any reasonable accuracy in these series of compounds we are not in a position to comment further. 

\section{Summary and conclusions}
The present results of the spontaneous magnetoelastic effects on CrF$_2$ and CuF$_2$ along with our previous similar results \cite{chatterji10a}  on CoF$_2$ and also the results\cite{chatterji10b} on MnF$_2$, FeF$_2$ and NiF$_2$  essentially complete this experimental investigation of the whole iron-group transition metal difluoride series except for VF$_2$ that has a very different spiral magnetic structure. We have detected and measured considerable exchange striction in each of these transition metal difluorides. We have established that the lattice strains couple with the square of the order parameter. However no systematic correlation between the magnetic properties and the magnetoelastic effects has been determined so far. The effect seems to increase with the magnetic moment value but even this correlation is not perfect. At this stage it seems necessary to do first-principle calculations and compare the results with the experimental ones. This may lead to the understanding of magnetoelastic effect in these simple insulating transition metal difluoride antiferromagnetic model system.

\end{document}